\documentclass[12pt]{article}
\usepackage{amssymb}
\usepackage{color,graphicx}
\usepackage[dvipsnames]{xcolor}
\usepackage{amsmath}
\usepackage{array}
\usepackage{epsfig} 
\usepackage{enumerate}
\usepackage{comment}
\usepackage{multirow}
\usepackage{bm}

\newcommand{\btpsi}{\tilde{\psi}_0}
\newcommand{\tpsi}{\tilde{\psi}}
\newcommand{\bd}{{D}_0}
\newcommand{\btau}{{\tau}_0}
\newcommand{\bttau}{\widetilde{\tau}_{0}}
\newcommand{\ttau}{\widetilde{\tau}}
\newcommand{\tA}{\tilde{A}}
\newcommand{\bpsi}{{\psi}_0}
\newcommand{\bg}{{g}_0}
\newcommand{\bmm}{{m}_0}
\newcommand{\bphi}{{\varphi}_0}

\newcommand{\btphi}{{\tilde{\varphi}}_0}
\newcommand{\bh}{{h}_0}
\newcommand{\bu}{{u}_0}
\newcommand{\tgamma}{\widetilde{\Gamma}}

\newcommand{\mx}{ \bm x }
\def\veps{{\varepsilon}}
\def\boldnabla{{\bm \nabla}}

\newcommand{\mk}{ \bm k }
\newcommand{\mpp}{ \bm p }

\def\dRM{\mathrm{d}}

\newcommand{\SA}{ {\mathcal S} }
\newcommand{\WA}{ {\mathcal W} }
\newcommand{\Dcal}{ {\mathcal D} }

\usepackage{xcolor}
\usepackage{comment}

\allowdisplaybreaks
\allowbreak

\begin{document}

\title{Renormalization group analysis of directed percolation process: \\ 
Towards multiloop calculation of scaling functions}

\author{ Michal Hnati\v{c}$^{1,2,3}$, Matej Kecer$^1$,
  Tom\'{a}\v{s} Lu\v{c}ivjansk\'{y}$^{2}$,
  Luk\'a\v{s} Mi\v{z}i\v{s}in$^1$
  }

\maketitle              
\mbox{ }\\
$^1$ Joint Institute for Nuclear Research, 141980 Dubna, Russia \\
$^2$ Institute of Physics, Faculty of Science, P. J. Šafárik University, Park Angelinum 9, 040 01 Košice, Slovakia\\
$^3$ Institute of Experimental Physics, Slovak Academy of Sciences, Watsonova 47, 040 01 Košice, Slovakia\\

\begin{abstract}
In this work, we employ a field-theoretic renormalization group approach to
study a paradigmatic model of directed percolation. We focus on the perturbative
calculation of the equation of state, extending the analysis to the three-loop order in
the expansion parameter $\veps = 4 - d$. 
We show that a large group of the necessary three-loop Feynman diagrams 
can be mapped onto already existing three-loop results, and develop a 
technique for the calculation of the remaining - truly novel - ones. The described semi-analytic procedure is further used to verify existing two-loop results. The main aim of this study is to provide an update on this ongoing work, as full three-loop calculations utilizing the described procedure are in progress.

\end{abstract}
\section{Introduction}
Nonequilibrium phase transitions \cite{Hinrichsen2000,HHL2008,Tauber2014} constitute a well-known branch of statistical physics. 
Similar to static phenomena, in such systems, interactions between interacting
degrees of freedom lead to cooperative behavior on large spatio-temporal scales. Such intriguing behavior is located in the vicinity of a critical region, in which strong
  fluctuations of the order parameter are especially pronounced. 
In contrast to static phenomena, phase transitions in these systems 
are of a dynamic nature, and this brings about several notable, distinctive features. 
One of the most relevant
is related to the divergence of correlation length not only in the spatial directions, but 
in the time direction as well.
The system's nonequilibrium character can arise from various mechanisms. For instance, in fully developed turbulence, this is achieved by a continuous supply of energy, which is transferred to the microscopic scales, where it is finally dissipated into heat.
 On the other hand, a viable way to drive a system into a nonequilibrium state is to
 explicitly break ergodicity by some internal mechanism. 
 This can be brought about, in particular, in those systems possessing
 so-called absorbing and active states. The former correspond to such states that the system cannot leave once entered.
 Within them, any dynamic activity ultimately stops, and the system does not evolve anymore. 
 The paradigmatic model belonging to this class is known as the directed bond percolation process (DP), which is presumably one of the simplest conceivable candidates and is also known in the literature as a simple epidemic process. 
 Its main
properties are summarized by the Janssen-Grassberger conjecture 
\cite{Tauber2014,Janssen2005}, which consists of four properties: (i) the existence of the
unique absorbing state, (ii) the positive one-component order
parameter, (iii) short-range interactions, and (iv) no presence of additional
specific properties. To date, this DP hypothesis has not been rigorously proven.

From a more general perspective, DP does not correspond to a concrete physical system, but rather it should be regarded as a prototypical representative of a broad universality class of systems. Any system fulfilling the aforementioned 
Janssen-Grassberger conjecture \cite{Janssen1981,Grassberger1982} is believed to belong to the DP universality class exhibiting the same critical behavior. 

Critical behavior is notorious for difficulties related to its direct
experimental verification.
In this regard, let us note that applications of DP universality have been found 
in research areas as diverse as high energy physics \cite{Moshe1978,Cardy1980},
 reaction-diffusion problems \cite{Odor2004}, a transition from
laminar to turbulent flow \cite{Lemoult2016,Sano2016,Hof2023}.

To study critical behavior in DP, several theoretical routes are available. For analytical treatment, a highly efficient approach is to apply field-theoretic methods. First, the DP is recast in terms of a functional integral. Second, relevant quantities 
(Green's functions) are expressed via a perturbation theory in the form of Feynman diagrams. Third, ultraviolet (UV) divergences present in Feynman diagrams are treated with a perturbative field-theoretic renormalization group (RG) method.
The latter method not only furnishes a conceptual formalism but also provides powerful and versatile algorithms for computing universal quantities, such as critical exponents, amplitudes,
and perturbative analysis of scaling functions. The governing parameter in the RG procedure is the upper critical dimension, commonly denoted as $d_c$.
Above $d_c$, fluctuations of the order parameter are 
 negligible
and the predictions based on the mean-field theory yield correct values for the critical exponents. Below $d_c$, fluctuations dominate the behavior of the
 system and RG methods allow a systematic treatment of the infrared divergences
 via a nontrivial relation \cite{Vasiliev2004} to UV divergences at the critical dimension $d_c$.

The DP process was analyzed to the two-loop order in \cite{Janssen1981}.
For more details, the reader is also referred to the review paper \cite{Janssen2005}, and related technical issues 
can be found in \cite{Janssen2001}.
Let us note that in general, perturbative calculations 
beyond the two lowest orders are not straightforward. Aside from
the large number of Feynman graphs to be evaluated, to 
determine the relevant divergent parts, we have to overcome nontrivial technical difficulties originating from the proper evaluation of divergent Feynman subgraphs.
 Perturbative calculations in general effective models lead
to only asymptotic series for relevant universal quantities. 
It is of imminent interest to determine as many loop corrections as 
feasible, since the subsequent resummation techniques \cite{Zinn2002} often yield very precise estimations comparable to Monte-Carlo predictions. 

Recently, we were able to accomplish semianalytic three-loop calculations for the DP process \cite{Adzhemyan2023}, and the isotropic percolation process as well \cite{Hnatic2025}.
The term semi-analytic refers to a specific way in which calculations were
performed. Specifically, we have worked analytically throughout, using numerical techniques to compute the arising integral.  In these papers, our main aim was the determination of fixed points of RG equations and the calculation of associated critical indices stemming from a linearization of RG flow equations in the vicinity of a given fixed point \cite{Vasiliev2004,Privman1991}.
However, it is well known that additional quantities, such as amplitude ratios and scaling functions, also exhibit universal behavior. 
It can be argued \cite{HHL2008,Privman1991} that the critical exponents are less sensible on the RG flow than scaling functions, since the latter receive contributions from the entire non-linear RG flow.
Based on this observation, it can be claimed that knowledge of scaling functions provides more convincing information on a studied universal behavior \cite{HHL2008,Lubeck2004} and thus provides more reliable validation of a given universality class. 

 One of the most studied scaling functions is related to the equation of state describing the behavior of the order
 parameter on a conjugated field.
  In magnetic systems, this corresponds to a relation between
 magnetization and the applied external magnetic field (also known in this context as the Widom-Griffiths scaling relation). For the DP process, this is given by a relation between the steady state value of infected individuals and, effectively, a probability for creating an infected individual locally at a given point, i.e., a measure of spontaneous creation of epidemic activity.
 The study of the corresponding equation of state was tackled by RG methods in the work \cite{Janssen1999}, and
our next impending goal is to further develop techniques and algorithms to extend three-loop calculations for such quantities. We believe that by pursuing this goal 
 for a relatively simple DP model, we will be able in the future to extend existing multi-loop calculations for more involved non-equilibrium models as well.

This paper is organized as follows. In Sec.~\ref{sec:model} we
provide essential information on the employed field-theoretic approach to the directed percolation problem. In Sec.~\ref{sec:diagram} we present 
 the main elements of the diagrammatic technique. Main steps of the RG approach are given in Sec.~\ref{sec:scaling_eqs}.
Concluding remarks are summarized in Sec.~\ref{sec:conclusion}.

{\section{Field-theoretic approach} 
\label{sec:model} }
In principle, there are two different ways for the derivation of field-theoretic
action for DP process~\cite{Tauber2014,Janssen2005}. Although they differ
in the final expressions, it can nevertheless
be shown that in the critical region they correspond to the
same critical theory. As this topic was already exposed in detail in the past, here, we refrain
from repeating the well-known derivation of the action and start directly
with the action functional.  We follow the nomenclature used in \cite{Janssen2005,Janssen1981} in which the formulation is based on the epidemiological parlance. This model can be interpreted
as a simplified model for epidemic spreading in which only two types of agents are present: susceptible and infected. The criticality appears at the border
between absorbing (no infected agents) and the active state (finite concentration of infected agents). The relevant
field variable is a coarse-grained density field $\psi_0=\psi_0(t,\mx)$, which corresponds to the local concentration of the infected agents.
As can be rigorously substantiated \cite{Tauber2014,Janssen2005}, the dynamic response functional for the DP process takes the abbreviated form
\begin{equation}
\SA = 
\btpsi \left( - \partial_t + \bd \boldnabla^2 - \bd \btau \right) \bpsi +\frac{\bd \bg}{2} \left(\btpsi^2 \bpsi -\btpsi \bpsi^2\right) + \bd \bh \btpsi ,
\label{eq:s_bare_DP}
\end{equation}
where the field $\bpsi \equiv \bpsi (t, \mx)$ is linked to the order parameters, $\btpsi \equiv \btpsi (t, \mx ) $ is the Martin-Siggia-Rose response field \cite{Tauber2014,Martin1973}, $\partial_t = \partial/ \partial t$ is the time derivative, $\boldnabla^2 = \sum_{i=1}^d \partial^2/ \partial x_i^2$ is the Laplace operator in $d$ spatial dimensions, $\bd$ is the diffusion constant, $g_0$ is the coupling constant, $\btau$ represents a deviation from the criticality, and $ \bd \bh \equiv \bd \bh (t, \mx)$ serves as an additional external source term. Let us note that whenever necessary, we also explicitly write the dependence of a given quantity on the space dimension $d$.
 This is motivated by the later use of the dimensional regularization method \cite{Vasiliev2004,Zinn2002} for an appropriate treatment of the ultraviolet divergences inherent to many Feynman diagrams. 
In the RG procedure, we have denoted all bare variables by a subscript "0". 
We have also introduced a condensed notation, in which integrals over the spatial variable $\mx$ and the time variable $t$ are implicitly assumed. For instance, the second term on the right-hand side of Eq.~\eqref{eq:s_bare_DP} corresponds to the integral expression
 $ \btpsi D_0 \boldnabla^2 \bpsi  = D_0 \int \dRM t\int \dRM^d x \,
  \btpsi (t,\mx) \boldnabla^2 \bpsi (t,\mx).$
Let us also point out that the functional formulation \eqref{eq:s_bare_DP}
corresponds to the stochastic differential equation (interpreted in the It\^o sense) of the form
\begin{equation*}  
   \partial_t \bpsi = D_0(\boldnabla^2 - \tau_0)\bpsi
   - \frac{D_0 g_0}{2} \bpsi^2 + D_0 h_0 + \sqrt{\bpsi} \zeta,
\end{equation*}
where the last term contains stochastic noise field $\zeta=\zeta(t,\mx)$ 
with the correlator of the form 
$\langle \zeta(t,\mx) \zeta(t',\mx') \rangle = g_0 D_0 \delta(t-t')
\delta^{(d)}(\mx-\mx')$.
This case is also known in the literature as a multiplicative noise.

In statistical physics, the typical aim of the theory is to determine
 behavior of correlation and response functions in the relevant asymptotic region
  of small frequencies/momenta (corresponding to large temporal and spatial scales). 
  
  The general framework for their definition is provided by the generating functional
\begin{equation}
\WA[\tA, A] = \ln \int \Dcal[\btpsi, \bpsi] \, \exp \left( \SA + \tA  \btpsi + A \bpsi \right),
\end{equation}
where  $A$ ($\tA$) is the source term associated with the field $\bpsi$ ($\btpsi$). The correlation functions, or Green functions, are obtained by taking functional derivatives of $W$ with respect to these sources. For translationally invariant systems, it is often advantageous to consider the irreducible Green functions, which correspond to the sum of all one-particle irreducible (1PI) Feynman diagrams in a perturbative expansion \cite{Zinn2002}. These diagrams are those diagrams that remain connected even if one internal line is cut.

The primary objective is to investigate the universal properties of the equation of state in the critical region $\tau_0\rightarrow 0$.
The universal behavior can be directly analyzed by a perturbative analysis of the Green functions of the DP model. In order to determine the equation of state, we first have to perform a shift in the bare response functional \eqref{eq:s_bare_DP} by incorporating a non-zero mean value for the order parameter, $\langle \bpsi \rangle \equiv \bmm \neq 0$. This can be conveniently achieved by replacing the original field $\bpsi$ with its mean value $\bmm$ and a new fluctuation field
$\bphi$ through the substitution $\bpsi = \bmm + \bphi$. 
After this shift, the response functional for DP becomes
\begin{equation}
\begin{aligned}
\SA &= 
\btphi \left( -\partial_t + \bd \boldnabla^2 - \bd \btau - \bd \bg \bmm \right) \bphi + \frac{\bg \bd}{2} \bmm {\btphi}^2 \\ 
&+ \frac{\bd \bg}{2} \left(\btphi^2 \bphi -\btphi \bphi^2\right) + \btphi \bd \left[ \bh - \bmm \left( \btau  +  \frac{1}{2} \bg \bmm \right) \right], 
\end{aligned}
\label{eq:s_bare_DP_shift}
\end{equation}
where, to have compact notation, we have relabeled the response field in the following way $\btphi \equiv \btpsi$.  

The equation of state in the mean-field (zeroth approximation) is obtained by minimizing the action functional with respect to the 
field $\btphi$, and neglecting 
fluctuations. To go beyond this approximation, it is necessary to compute corrections in a perturbative fashion. This involves, in particular, calculating the Green function $\langle \btphi \rangle$, which quantifies the system's response to the external field $\bd \bh$, around the shifted (i.e., mean-field) background. From these perturbative considerations, we can derive the following formal expression for the equation of state 
\begin{equation}
0 = \bd \bh - \bd \bmm \left(\btau + \frac{1}{2} \bg \bmm \right)  
+\sum\limits_\text{graphs},
\label{eq:pert_form_es}
\end{equation}
where the term $\sum\limits_\text{graphs}$ 
stands for all diagram corrections composed of 1PI Feynman diagrams with only one external $\btphi$-leg. The RG method can then be utilized to systematically treat the UV divergences that appear at the 
 upper critical dimension. Since the shifted model is logarithmic at the same space dimension $d_c = 4$ as the original DP model, the actual RG analysis is then akin to the fixed point analysis of the DP process \cite{Janssen2005,Adzhemyan2023}. 
 For convenience, we employ the dimensional regularization method \cite{Vasiliev2004,Collins1984}
 and introduce the formally small parameter $\veps$ as follows 
\begin{equation}
\veps = d_c-d = 4 - d.
\end{equation}
It corresponds to a deviation from the upper critical dimension of the 
model, and in perturbation theory, it serves two purposes.
First, the actual UV divergences manifest themselves in Green functions as poles in $\veps$ \cite{Vasiliev2004,Zinn2002}, and thus $\veps$ naturally provides
a parameter for a systematic treatment of multi-loop calculations.
Second, subsequent fixed-point analysis yields perturbative estimations
of various physical quantities in the form of asymptotic series in $\veps$.

\section{Diagrammatic analysis}
\label{sec:diagram}

\begin{figure}[!ht]
\centering
\includegraphics[width=0.75\linewidth]{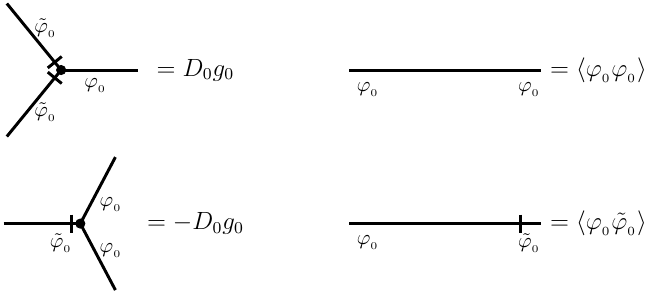}
\caption{The graphical representation of Feynman rules for the DP model
 obtained from the shifted action functional~\eqref{eq:s_bare_DP_shift}.}
\label{fig:feyn_rules}
\end{figure}

Feynman diagrams serve as a powerful graphical representation for the algebraic structures arising in perturbation theory \cite{Vasiliev2004,Zinn2002}.
These diagrams are constructed from fundamental perturbative elements, namely propagators and interaction vertices. For the action~\eqref{eq:s_bare_DP_shift}, the two bare propagators are derived from the quadratic terms of the shifted action functional \eqref{eq:s_bare_DP_shift}. In the frequency-momentum representation $(\omega,\mk)$, the bare propagators take the form
\begin{equation}
\langle \bphi \btphi \rangle = \langle \btphi \bphi \rangle^{*} =  \frac{1}{-i \omega + \bd(\mk^2 + \bttau)}, 
\quad
\langle \bphi \bphi \rangle =  \frac{\bd \bg \bmm}{\omega^2 + \bd^2(\mk^2 + \bttau)^2}
,
\end{equation}
where we have introduced the shifted temperature parameter $\bttau \equiv \btau + \bg \bmm$. In the time-momentum representation, the propagators take the form
\begin{equation}
\begin{split}
\langle \bphi(t) \btphi(t') \rangle &= \theta (t - t') \exp \{ 
-\bd(\mk^2 + \bttau) (t -t')\}, \\
\langle \bphi(t) \bphi(t') \rangle &= \frac{\bg \bmm}{2(\mk^2 + \bttau)}
\exp \{ -\bd(\mk^2 + \bttau)|t -t'| \}, 
\end{split}
\end{equation}
where  $\theta (t)$ is the Heaviside step function, and, for brevity, we have suppressed the $\mk$-dependence of quantities on the left-hand sides.
The nonlinear terms in the action functional \eqref{eq:s_bare_DP_shift} generate two interaction vertices of the model, which are linked in the perturbation theory with the following vertex factors \cite{Vasiliev2004}
\begin{equation}
V_{\bphi \btphi \btphi} = - V_{\bphi \bphi \btphi} = D_0 g_0   .
\end{equation}
It is easy to observe that the interaction part of the shifted response functional \eqref{eq:s_bare_DP_shift} takes the identical form as that in the original model \eqref{eq:s_bare_DP}. Finally, we are thus able to summarize all needed Feynman rules for the shifted DP model in the compact graphical representation (see~Fig.~\ref{fig:feyn_rules} for the chosen depiction). 

Let us further analyze the 1PI Feynman diagram with a single $\btphi$-leg, containing an arbitrary number of loops $l$, the number of propagators $p$, and vertices $v$. It is possible to express the number of propagators (vertices) in such a diagram in terms of the loops $l$ as
\begin{equation}
p = 3 l - 2, \hspace{2cm} v = 2 l - 1.
\label{eq:number_prop_vert_loop}
\end{equation}
It is important to note that there are two different types of propagators and also vertices in the shifted model. The total number of propagators is $p = p_1 + p_2$, where $p_1$ denotes the number of $\langle \bphi \btphi \rangle$ propagators and $p_2$ the number of $\langle \bphi \bphi \rangle$ propagators, respectively.
Similarly, the total number of interaction vertices is separated into the sum $v = v_1 + v_2$, where $v_1$ counts $\langle \bphi \btphi \btphi \rangle$ vertices, whereas $v_2$ counts $\langle \bphi \bphi \btphi \rangle$ vertices.
From the equality of the number of fields $\bphi$, $\btphi$ at the vertices and propagators, the following relation can be easily derived
\begin{equation}
\begin{split}
\textrm{Field } \bphi : \hspace{0.5cm} & v_1 + 2 v_2 = p_1 + 2 p_2, \\
\textrm{Field } \btphi : \hspace{0.5cm} & 2 v_1 + v_2 = p_1 + 1.
\end{split}
\label{eq:number_fields}
\end{equation}
Combining relations (\ref{eq:number_prop_vert_loop}) and (\ref{eq:number_fields}) leads to the following formulae
\begin{equation}
l = v_1 + p_2, \quad
l + p_2 = v_2 + 1 .
\end{equation}
The first relation implies that the number of $\langle \bphi \bphi \rangle$ propagators $p_2$ cannot exceed the loop number $l$. The maximum $p_2 = l$ occurs for $v_1 = 0$. 
Therefore, with the maximum number of propagators $p_2$, the diagram is composed solely of vertices $\langle \bphi \bphi \btphi \rangle$. The minimum is achieved for $p_2 = 1$; otherwise, the diagram would contain a closed loop consisting of retarded $\langle \bphi \btphi \rangle$ propagators.

\begin{figure}[!ht]
\centering
\includegraphics[width=1.0\linewidth]{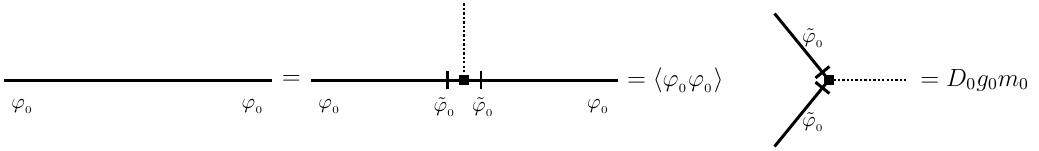}
\caption{The graphical representation of the propagator $\langle \bphi \bphi \rangle$ and the 
newly generated interaction vertex.}
\label{fig:propagator_pp}
\end{figure}

The quadratic part of the bare action functional \eqref{eq:s_bare_DP_shift} indicates that there are two propagators with a structure similar to the model A of critical dynamics \cite{Tauber2014,Vasiliev2004}. Following similar considerations as applied in the analysis of this model A, the shifted model for DP offers an opportunity to express the propagator $\langle \bphi \bphi \rangle$ in terms
$\langle \bphi \btphi \rangle$ through the relation
\begin{equation}
\langle \bphi \bphi \rangle = \bd \bg \bmm \langle \bphi \btphi\rangle \langle\btphi \bphi \rangle.    
\end{equation}
 This relation, depicted schematically in Fig.~\ref{fig:propagator_pp}, displays a new representation for the propagator $\langle \bphi \bphi \rangle$ and effectively introduces
 a new vertex with the algebraic factor $\bd \bg \bmm$.
  As it will become more transparent later on through analysis of specific diagrams, this construction is essential for linking the diagrammatic technique of shifted functional \eqref{eq:s_bare_DP_shift} to that of the original model \cite{Adzhemyan2023}. All that is needed is to connect the external leg of the new vertex to the
 field $\bphi$.

The corrections to the equation of state arising from 1PI diagrams can be systematically organized into the classes distinguished by the number of loops and number of
propagators $\langle \bphi \bphi \rangle$. In general, the diagrammatic contributions in Eq.~\eqref{eq:pert_form_es} can be formally expressed as the sum
\begin{equation}
\sum\limits_\text{graphs}
= \sum\limits_{n = 1}^{l} (\bd \bg)^{2n-1} \sum\limits_{j = 1}^{n} (-1)^{n+j-1}(\bd \bg \bmm)^{j} \, \Gamma_{n, j} (\bd, \bttau, \veps),
\label{eq:graphs_expansion}
\end{equation}
where the common factor 
$\bd \bg \bmm$ has been extracted from each $\langle \bphi \bphi \rangle$ propagator. In the following considerations, we will understand by propagators only the part without the vertex factor.

To first order in perturbation theory, only one diagram contributes, which can be graphically represented 
in the following way
\begin{equation}
\raisebox{-10.mm}{\includegraphics[width=8.5cm]{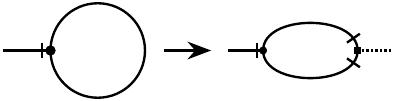}} 
= -\bd^2 \bg^2 \bmm \, \Gamma_{1,1} (\bd, \bttau, \veps)
.
\end{equation}
After integration over the internal frequency and momentum, we get
\begin{equation}
\Gamma_{1,1} (\bd, \bttau, \veps) = \frac{\Gamma \left(1+\frac{\veps}{2} \right)}{(4\pi)^{d/2}} \frac{\bttau^{1-\veps / 2}}{\bd} \frac{1}{\veps (\veps-2)} .
\end{equation}
Hence, at first sight novel one-point diagram in fact corresponds to the appropriate self-energy diagram of the original DP model, yielding the same contribution. As we will show, such a property significantly reduces the number of Feynman diagrams needed for calculations in higher orders in perturbation theory.

In general, 
it holds that any Feynman diagram containing one propagator $\langle \bphi \bphi \rangle$ is related to a self-energy diagram 
of the original model \eqref{eq:s_bare_DP}. 
Such contributions were computed up to the third order in \cite{Adzhemyan2023}, where the information was obtained from diagram $\langle \btphi \bphi\rangle$ using the following formula
\begin{equation}
\tgamma (\btau, \veps) = - 
\frac{\partial}{\partial \btau} 
\langle \btphi \bphi \rangle \bigg|_{\mpp = 0, \, \omega =0}.
\end{equation}
Here, the $\tgamma (\btau, \veps)$ stands for the contribution of any 
self-energy diagram to the RG constant. The 
vertex factors $V_{\bphi \bphi \btphi}$ and $V_{\bphi \btphi \btphi}$, and parameters such as $\bg$ and $\bd$ are omitted in this relation, and therefore $\tgamma$ depends only on parameters
$\btau$ and $\veps$. However, $\tgamma (\btau,\veps)$ 
can be factorized as $\btau^{-n\veps/2}$, where $n$ is the number of loops, leaving a purely $\veps$-dependent Laurent series $\tgamma (\veps)$.

In order to utilize the results of \cite{Adzhemyan2023}, it is thus necessary to express the diagram contribution using the derivative with respect to $\bttau$. 
This can be readily accomplished;
consider a $n$-loop diagram in frequency-momentum representation. After integration over internal frequency $\omega_{\mk}$ and rescaling
an internal momentum $\mk$ according to the prescription $\mk \rightarrow \bttau^{1/2} \mk$, the final contribution 
takes the following form
\begin{equation}
\Gamma_{n,1}^{(i)} (\bd, \bttau, \veps) = \frac{\bttau^{1 - \veps n / 2}}{\bd^{2n-1}} \Gamma_{n,1}^{(i)} (\bttau = 1, \veps),
\label{eq:self_ene_tau_1}
\end{equation}
where $\Gamma_{n,1}^{(i)}$ denotes 
the $i$-th Feynman diagram in $n$-loop approximation with
one $\langle \bphi \bphi \rangle$ propagator. Applying the differential operator $ \bttau \partial_{\bttau}$ on 
the Eq.~\eqref{eq:self_ene_tau_1} and rearranging all parameters yields
\begin{equation}
\Gamma_{n,1}^{(i)} (\bd, \bttau, \veps) = \frac{ \bttau}{1 - \veps \, n /2} \frac{\partial}{\partial \bttau} \Gamma_{n,1}^{(i)} (\bd, \bttau, \veps).
\end{equation}
Since the control parameter $\btau$ plays the same role in integrals (and hence in derivatives) as $\bttau$. Therefore, the calculation of contribution of diagram $\Gamma_{n,1}^{(i)}$ can be expressed using the corresponding self-energy diagram as follows
\begin{equation}
\Gamma_{n,1}^{(i)} (\bd, \bttau, \veps) = \frac{- 1}{1 - \veps \, n / 2} \frac{s_\text{shift}}{s_\text{orig}} \frac{\bttau} {\bd^{2n-1}} 
\tgamma (\btau \equiv \bttau, \veps)
= \frac{- 1}{1 - \veps \, n / 2} \frac{s_\text{shift}}{s_\text{orig}} \frac{\bttau^{1-\veps n/2}}{\bd^{2n-1}} \tgamma (\veps),
\label{eq:reation_phiphi_1}
\end{equation}
where the $s_\text{shift}$ corresponds to the symmetry factor of 
 1PI $\langle \btphi \rangle$ diagram and $s_\text{orig}$ is symmetry factor of corresponding self-energy 
diagram $\langle \btphi \bphi\rangle$. 
In addition, all dependencies on $\bd$ and $\bttau$ have been factored out. 

\begin{figure}[!ht]
\centering
\includegraphics[width=0.48\linewidth]{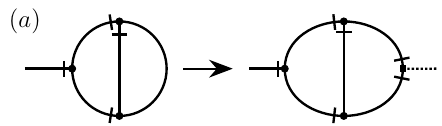} \hspace{3mm}
\includegraphics[width=0.48\linewidth]{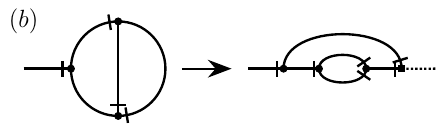}
\newline \\
\includegraphics[width=0.48\linewidth]{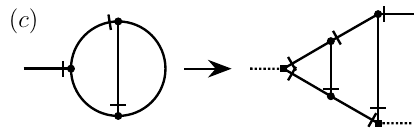} \hspace{3mm}
\includegraphics[width=0.48\linewidth]{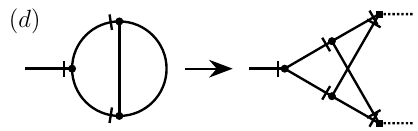}
\caption{The graphical representation of the two-loop diagrams with the symmetry coefficient: (a) $s_\text{shift} = 1$ (b) $s_\text{shift} = 1/2$ (c) 
$s_\text{shift} = 1$ (d) $s_\text{shift} = 1/4$.}
\label{fig:2_loop_diagrams}
\end{figure}

Another class of 1PI Feynman diagrams whose contributions can be expressed in terms of the original model's diagrams is that containing exactly two $\langle \bphi \bphi \rangle$ propagators. More precisely, we refer here to the $\langle \btphi \bphi^2 \rangle$-type vertex diagrams, examples of which are shown in Fig.~\ref{fig:2_loop_diagrams}~(c)~--~(d). In this case, the situation is simpler, and the contribution of the vertex diagram itself takes the form \cite{Adzhemyan2023}
\begin{equation}
\tgamma (\btau, \veps) = -  
\langle \btphi \bphi^2 \rangle \bigg|_{\mpp = 0, \,  \omega =0}
,
\end{equation}
where the overall minus sign stems from the normalization of the vertex function $\langle \btphi \bphi^2 \rangle$, and parameters such as $\bg$ and $\bd$ are, for brevity, omitted in this relation. Employing a procedure analogous to the previous case, the contribution of the diagram $\Gamma^{(i)}_{n,2}$ can be represented in terms of the corresponding vertex diagram as follows
\begin{equation}
\Gamma^{(i)}_{n,2} (\bd,\bttau, \veps) = \frac{\bttau^{-n \veps / 2}}{\bd^{2n}} \Gamma^{(i)}_{n,2} (\bttau = 1, \veps) = - \frac{s_{shift}}{s_{orig}} \frac{\bttau^{-n \veps / 2}}{\bd^{2n}} 
\tgamma (\veps),
\label{eq:reation_phiphi_2}
\end{equation}
where $\tgamma (\veps)$ denotes $\veps$-dependent Laurent series of corresponding vertex diagrams. 

Consequently, utilizing the derived relations ~\eqref{eq:reation_phiphi_1} and \eqref{eq:reation_phiphi_2}, a full two-loop approximation to the equation of state can be expressed solely through already existing results \cite{Adzhemyan2023}. 
Namely, to second order in perturbation theory, there are four two-loop diagrams, which are all given in Fig.~\ref{fig:2_loop_diagrams}. 
They are naturally 
 divided into two distinct classes, the first class 
  corresponds to the
 two diagrams with one $\langle \bphi \bphi \rangle$, 
 whereas the second class 
 contains two diagrams with two $\langle \bphi \bphi \rangle$ propagators.
  Their respective contributions read
\begin{align}
\Gamma^{(a)}_{2,1} & = - \frac{\bttau^{1-\veps}}{8\bd^{3}} \frac{\Gamma(1+\veps)}{(4\pi)^d \veps (1-\veps)} \left[ \frac{4}{\veps} + 2 - 2.988841305 
\veps + \mathcal{O}(\veps^2)\right], \label{eq:gamma_a21}\\
\Gamma^{(b)}_{2,1} & = \frac{\bttau^{1-\veps}}{\bd^{3}} \frac{\Gamma(1+\frac{\veps}{2})}{(4 \pi)^{d} \Gamma(2 - \frac{\veps}{2})} \frac{3}{16 \veps} \left[1 + \left(2 - \ln 3\right) \frac{\veps}{2} + \mathcal{O}(\veps^2) \right]
,\\
\Gamma^{(c)}_{2,2} & = \frac{\bttau^{-\veps}}{\bd^4} \frac{\Gamma(1+\veps)}{16 (4\pi)^d \veps} \left[\frac{2}{\veps} + 1 - 6 \ln 2 + 3 \ln 3 - 0.816060873   \veps +  \mathcal{O}(\veps^2)\right],\\
\Gamma^{(d)}_{2,2} & = \frac{\bttau^{-\veps}}{\bd^4} \frac{\Gamma (1+\veps)}{(4\pi)^d \veps} \frac{3}{64} \left[ 4 \ln \frac{4}{3} - 0.904479706  \veps+ \mathcal{O}(\veps^2)  \right],
\label{eq:two_loop_result}    
\end{align}
where all numerical values with decimal places have been rounded, and 
$\Gamma (x)$ denotes Euler's gamma function.
To compare the results, it is useful to introduce a new coupling constant $\bu$ 
  through the following substitution
\begin{equation}
 \bg^2 \frac{S_d}{2 (2\pi)^d} \, \rightarrow \, \bu , \hspace{1cm}
 S_d = \frac{2\pi^{d/2}}{\Gamma(d/2)}
 ,
\end{equation}
where the latter equation gives the surface of the unit sphere $S_d$ in the $d$-dimensional space. We effectively absorb the common geometric factor commonly arising in the computation of Feynman diagrams to all orders in perturbation theory. Then, Eq.~\eqref{eq:pert_form_es} can be rewritten up to second-order in perturbation theory as follows:
\begin{equation}
\bd \bh = \bd \bmm \left(\btau + \frac{\bg \bmm }{2}\right) - \bd \bmm \left[
-\bu \bttau^{1-\veps/2} \Gamma_{1,1} +
\bu^2 \bttau^{-\veps} \left(  \bttau \Gamma_{2,1} - \bg \bmm \Gamma_{2,2} \right) \right],
\label{eq:state_2_loop}
\end{equation}
where $\Gamma_{i,j}$ represents a Laurent series in parameter
$\veps$. The two-loop contributions to the coefficients are listed in Table~\ref{tab:diagram_2_loop}, where we have used the following abbreviations
$\Gamma_{2,1}^{(a)} + \Gamma_{2,1}^{(b)} \equiv \Gamma_{2,1} $ and $\Gamma_{2,2}^{(c)} + \Gamma_{2,2}^{(d)} \equiv \Gamma_{2,2} $. 

\begin{table}[!ht]
\renewcommand{\arraystretch}{2.0}
\centering
\begin{tabular}{c|c|c|l|l}
\hline
$\Gamma_{i,j}^{(\dots)}$ & method & $\veps^{-2}$ & $\veps^{-1}$ & $\veps^{0}$ \\
\hline
\hline
\multirow{2}{*}{$\Gamma_{2,1}^{(a)}$} & exact & $-\frac{1}{2}$ & $-0.25$ & $-0.3682451^{*}$\\
& numeric & $-\frac{1}{2}$ & $-0.250000(2)$ & $ -0.368239(6)$\\
\hline
\multirow{2}{*}{$\Gamma_{2,1}^{(b)}$} & exact & $0$ & $0.1875$ & $-0.009 244 9^{*}$ \\
& numeric & $0$ & $0.1875000(1)$ & $ -0.009244(10)$ \\
\hline
\multirow{2}{*}{$\Gamma_{2,2}^{(c)}$} & exact & $\frac{1}{8}$ & $-0.1164404^{*}$ & $0.1258992^{*}$\\
& numeric & $\frac{1}{8}$ & $-0.116441(4)$ & $0.125901(4)$\\
\hline
\multirow{2}{*}{$\Gamma_{2,2}^{(d)}$} & exact & $0$ & $0.0539404^{*}$ & $-0.09633788^{*}$ \\
& numeric & $0$ & $0.053940(1)$ & $-0.0963372(12)$ \\
\hline
\end{tabular}
\caption{The coefficients for Feynman diagrams in two-loop approximation. The star symbol~$^{*}$ indicates that the exact value is rounded and the number in brackets corresponds to the numerical error in the last digits. }
\label{tab:diagram_2_loop}
\end{table}

At three-loop order, the total contribution comprises $65$ Feynman diagrams. These diagrams can be categorized by the number of $\langle \bphi \bphi \rangle$ propagators. The first group, containing $17$ diagrams, features a single such propagator. 
Their contributions can thus be computed using relation \eqref{eq:reation_phiphi_1} by determining the corresponding self-energy diagram. There are $32$ diagrams in the second group, all of them containing two instances of $\langle \bphi \bphi \rangle$ propagator. Their contributions are calculated via relation \eqref{eq:reation_phiphi_2} respectively, using the corresponding vertex diagram. Therefore, from the total number of 65 three-loop diagrams, 49 can be readily obtained from already existing results \cite{Adzhemyan2023}.
{It is only the remaining third group 
formed by diagrams containing three $\langle \bphi \bphi \rangle$ propagators which leads to the
truly novel contributions and have
to be computed in full. Unlike the previous groups, the contributions from these $16$ diagrams cannot be expressed in terms of diagrams of the original model~\eqref{eq:s_bare_DP}. 

To obtain these, a numerical method similar to that of \cite{Adzhemyan2023} was developed, integrating the Feynman integrals via Sector Decomposition \cite{Binoth2000} and evaluating them numerically with the Vegas algorithm from the Cuba library \cite{Hahn2005}. Software implementation of these algorithms for the particular case of appearing one-point diagram expressions was completed and further tested on two-loop diagrams. In this case, the obtained values can be directly compared to analytical results \eqref{eq:gamma_a21}-\eqref{eq:two_loop_result}. The corresponding numerical values are stated in Fig. \ref{fig:2_loop_diagrams} alongside their analytical counterparts, and demonstrate an excellent numerical accuracy 
 of the employed method.

The calculation of novel three-loop diagrams using the developed technique is ongoing.



\section{Scaling formulation of the equation of state}
\label{sec:scaling_eqs}

Computing the contribution of Feynman diagrams is only the first step in the procedure. To obtain the scaling form of the equation of state near the phase transition, the relation (\ref{eq:state_2_loop}) has to be rewritten in terms of renormalized quantities. The field shift does not affect the calculation of the renormalized quantities. Consequently, the renormalized action functional of the original model is sufficient and takes the abbreviated form:
\begin{equation}
\SA_R  = \tpsi \left( -Z_1\partial_t + Z_2 D \boldnabla^2 - Z_3 D\tau \right)
\psi + 
\frac{ D g \mu^{\veps/2}}{2} Z_4 \left(\tpsi^2 \psi -\tpsi \psi^2\right) + D h \tpsi
.
\end{equation}
The transition from bare $Q_0$ to renormalized quantities $Q$ is achieved straightforwardly through
 the substitution
\begin{equation}
Q_0 \rightarrow Z_Q Q,
\end{equation}
where $Q$ runs over all fields and parameters of the model $\{\tpsi,\psi,D,\tau,g,h \}$.
Relations between the renormalization constants of fields and parameters then directly follow, and read
\begin{equation}
\begin{split}
Z_h = Z_1^{1/2} Z_{2}^{-1}, \hspace{1cm} 
Z_D = Z_1^{-1} Z_2, \hspace{1cm}
Z_{\tau} = Z_2^{-1} Z_3, \\
Z_m = Z_{\psi} = Z_{\tpsi} = Z_1^{1/2}, \hspace{1cm}
Z_u = Z_g^2 = Z_1^{-1} Z_2^{-2} Z_4^{2}. 
\end{split}
\label{eq:RG_constant}
\end{equation}
The full information on RG constants $Z_i; i=1,2,3,4$ to the three-loop precision can be found in the article \cite{Adzhemyan2023}. By inserting the renormalized quantities into relation \eqref{eq:state_2_loop}, truncating the pole contributions of diagrams, we arrive at the following relation 
\begin{equation}
h  = m \left[\tau + \frac{g m}{2} + \tau A_1(g, \veps, \ln (\ttau / \mu^2) ) + g m A_2(g, \veps, \ln (\ttau / \mu^2))  \right], 
\label{eq:of_state_renorm}
\end{equation}
where functions $A_1$ and $A_2$ are UV finite, since they do not contain poles in $\veps$.
In what follows, using the result of RG analysis, this relation will be recast into the scaling form that 
 is valid near the phase transition point.



The main task at this point is to construct the Widom-Griffiths scaling form of the equation of state near the transition. For the DP model, the construction starts from the RG equation in bare parameters   
\begin{equation}
\mu \frac{d}{d \mu}\bigg|_0 \bh \left( \btau,  \bmm, \bg \right) = 0
.
\end{equation}
After introducing renormalized parameters, the RG equations can be solved to obtain the RG functions (anomalous dimensions and beta function $\beta(g)$). From the condition $\beta(g) = 0$, the coordinate $g \equiv g^{*}$ of the infrared fixed point is determined.  Combining these RG functions with dimensional analysis yields the following scaling law
\begin{equation}
h \left(\tau, m, g; \mu \right) = \mu^{2 + d/2} s^{d +\Delta_{\omega}-\Delta_{\varphi}} h \left(\mu^{-2} s^{-\Delta_{\tau}} \tau, \mu^{-d/2} s^{-\Delta_{\varphi}} m, g_{*};1\right)
,
\end{equation}
where $s$ denotes the flow parameter. Critical dimensions $\Delta_{\varphi}, \Delta_{\omega}$, and $\Delta_{\tau}$ are determined from the RG constants \eqref{eq:RG_constant} and have been calculated up to three-loop order recently~\cite{Adzhemyan2023}.

To investigate the equation of state in the critical region $( \tau, m \rightarrow 0)$, it is necessary to map this region by an appropriate choice of flow parameter $s$ onto scales, on which perturbation theory can be justifiably applied:
\begin{enumerate}[(1)]    
\item The flow parameter chosen according to $s = m^{1 / \Delta_{\varphi}}$ leads to the Widom-Griffiths scaling form 
\begin{equation}
  h = m^{(d + \Delta_{\omega})/ \Delta_{\varphi} -1} 
  F(\tau m^{-  \Delta_{\tau} / \Delta_{\varphi} }).
\end{equation}
\item On the other hand, the flow parameter $s = \tau^{1 / \Delta_{\tau}}$ yields the scaling form 
\begin{equation}
h = \tau^{(d + \Delta_{\omega} -\Delta_{\varphi}) / \Delta_{\tau}} F(m \tau^{-\Delta_{\varphi} / \Delta_{\tau} }).
\end{equation}
\end{enumerate}

These two possible scaling forms of the equation of state in the critical region provide a path to obtain a universal amplitude combination. This combination is derived by expressing Eq.~\eqref{eq:of_state_renorm} in the following scaling form:
\begin{equation}
 a(\veps) h = m^{(d + \Delta_{\omega})/ \Delta_{\varphi} -1} F \left( b(\veps) \tau m^{-\Delta_{\tau} / \Delta_{\varphi}}\right). 
\end{equation}
The construction proceeds order by order in perturbation theory, with scaling variables $x = b(\veps) \tau m^{-\Delta_{\tau} / \Delta_{\varphi}}$ and $y = a(\veps) h m^{1 - (d + \Delta_{\omega})/ \Delta_{\varphi}} $ introduced. To construct the universal scaling function, we invoke the normalization condition 
\begin{equation}
F (0) = 1, \hspace{1cm} F(-1) = 0.
\end{equation}
The non-universal features are absorbed into two non-universal parameters $a(\veps), b(\veps)$ leading to the universal scaling function. In this construction, the universal scaling function $F(x)$ is identical for all systems belonging to the given universal class \cite{Lubeck2004}. Further details of this construction at the two-loop order are available in the literature \cite{Janssen1999}, 
and we do not repeat them here.


Another universal quantity that can be extracted from this scaling form of the equation of state is a so-called amplitude ratio \cite{Privman1991}. 
More precisely, we consider the amplitude ratio associated with the susceptibility
 in the vicinity of the phase transition. In the DP model, the susceptibility $\chi$ is defined 
 in an expected way~\cite{HHL2008,Janssen1999} as the derivative of the order parameter $m$ with respect to the external source $h$:
\begin{equation}
\chi = \frac{\partial m}{\partial h} \bigg|_{\tau} \,.    
\end{equation}
Its determination requires switching to a parametric representation valid over the entire $(\tau, m)$ domain. Applying this procedure, as detailed in \cite{Janssen1999}, yields the following result for the universal amplitude ratio at the two-loop level:
\begin{equation}
\frac{\chi_{-}}{\chi_{+}} = \frac{2\Delta_{\varphi}}{\Delta_{\tau}} - 1 + \mathcal{O}(\veps^3)  = 1 - \frac{\veps}{3} - \veps^2 0.022 563 \dots \, + 
\mathcal{O}(\veps^3),
\label{eq:amplitude_ratio}
\end{equation}
where $\chi_{-}$ ($\chi_+$) correspond to the limit case $\tau \rightarrow 0^{-}$ 
($\tau \rightarrow 0^+$).
It remains an open question whether the corresponding relation \eqref{eq:amplitude_ratio} will be corrected at the third (higher) order in perturbation theory or not. 
Moreover, the universal amplitude ratio can be compared with results obtained from lattice models of directed percolation \cite{Lubeck2004}. Consequently, we can evaluate the extent to which this universal ratio is modified by higher-order contributions. As argued in \cite{Janssen1999}, this can be interesting from the practical point of view.

{ \section{Conclusion} \label{sec:conclusion} }

In this paper, we present recent progress in multiloop perturbation theory for universal properties in nonequilibrium systems, specifically for the directed percolation class. This study computes the scaling form of the equation of state in the critical region. We introduce a technique that uses diagram contributions from renormalization constants \cite{Adzhemyan2023} to evaluate a class of three-loop diagrams, thereby reducing the number of diagrams requiring direct computation from $65$ to $16$ (those with three $\langle \bphi \bphi \rangle$ propagators). Moreover, we develop a software for such direct computation of the remaining diagrams. At the two-loop level, we benchmark our method by comparing analytical and numerical results, finding excellent precision. This established numerical method will be applied to the full three-loop calculation.

Future work could extend this analysis in several directions. The most immediate task is to derive universal amplitude ratios and the universal scaling function $F(x)$ to $\mathcal{O}(\epsilon^3)$.  These universal quantities can be compared with independent studies based on Monte-Carlo simulations of various lattice models \cite{Lubeck2004} belonging to the DP universality class. A more demanding long-term goal is a complete four-loop calculation of the DP process, which is necessary for even higher-accuracy asymptotic scaling predictions. This is an important task also from a methodological point of view, since it is probable that the developed numerical techniques can be useful also for multiloop calculations of other models in nonequilibrium physics.


\section*{Acknowledgment}
The work was supported by VEGA Grant No. 1/0297/25
of the Ministry of Education, Science, Research and Sport
of the Slovak Republic. The authors have greatly benefited from
mutual collaborations and many valuable discussions with
Loran Ts. Adzhemyan.

Conflict of Interest: The authors declare that they have no conflicts of interest.


\begin{thebibliography}{99}  
\bibitem{Hinrichsen2000} H. Hinrichsen, Adv. Phys. {\bf 49},  815-958 (2000).

\bibitem{HHL2008} M. Henkel, H. Hinrichsen, and S. L\"ubeck, Non-
equilibrium phase transitions, vol. 1 (Springer, 2008).

\bibitem{Tauber2014} 
U. C.~T{\"a}uber, {\it Critical Dynamics: A Field Theory Approach to Equilibrium and Non-Equilibrium Scaling Behavior}, Cambridge University Press, Cambridge (2014).

\bibitem{Janssen2005}
H. K. Janssen, U. C. T\"{a}uber, Ann. Phys. {\bf 315}, 147 -- 192 (2005). 

\bibitem{Janssen1981} H.-K.~Janssen, Z. Phys. B {\bf 42}, 151 (1981).

\bibitem{Grassberger1982} P.~Grassberger, Z. Phys. B {\bf 47}, 365–374 (1982).

\bibitem{Moshe1978} M. Moshe, Phys. Rep. {\bf 37}, 255 (1978).

\bibitem{Cardy1980} J. L. Cardy and R. L. Sugar, J. Phys. A: Math. Gen. 
        {\bf 13}, L423 (1980).
        
\bibitem{Odor2004} G. \'Odor, Rev. Mod. Phys. {\bf 76}, 663 (2004).

\bibitem{Lemoult2016} G. Lemoult, L. Shi, K. Avila, S. V. Jalikop, M. Avila, and B.
Hof, Nat. Phys. {\bf 12}, 254 (2016).

\bibitem{Sano2016}  M. Sano and K. Tamai, Nat. Phys. {\bf 12}, 249 (2016).

\bibitem{Hof2023} B. Hof, Nat. Rev. Phys. {\bf 5}, 62-72 (2023).

\bibitem{Vasiliev2004} 
A. N. Vasil'ev, {\it The Field Theoretic Renormalization Group  in Critical Behavior Theory and Stochastic Dynamics}, [in Russian], PIYaF, St. Petersburg (1998); English trans., Chapman and Hall/CRC, Boca Raton, Fla (2004).

\bibitem{Janssen2001} H.-K.~Janssen, J. Stat. Phys. {\bf 103}, 801 (2001).

\bibitem{Zinn2002} 
J. Zinn-Justin, {\it Quantum Field Theory and Critical Phenomena}, Clarendon Press 1989, 4th  edn. Oxford University Press (2002).

\bibitem{Adzhemyan2023}
L. T. Adzhemyan, M. Hnati\v{c}, E. V. Ivanova, M. V. Kompaniets, T. Lu\v{c}ivjansk\'{y}, L.~Mi\v{z}i\v{s}in, Phys. Rev. E {\bf 107} (6), 064138 (2023).

\bibitem{Hnatic2025} M.~Hnati\v{c}, M.~Kecer, M.~V.~Kompaniets, 
T. Lu\v{c}ivjansk\'y, L.~Mi\v{z}i\v{s}in , Y.~G.~Molotkov, Phys. Rev. E {\bf 112}, 014113 (2025).

\bibitem{Privman1991} V. Privman, P. Hohenberg, and A. Aharony,
  {\it Phase transitions and critical phenomena} (Volume 14),
  Academic, New-York (1991).


\bibitem{Lubeck2004} S. L\"ubeck, Int. J. Mod. Phys. B {\bf 18}, 3977-4118 (2004).
  
\bibitem{Janssen1999}
H. K. Janssen,  \"{U}. Kutbay, K. Oerding, J. Phys. A: Math. Gen. {\bf 32}, 1809 -- 1817 (1999).

\bibitem{Martin1973}
P. C. Martin, E. D. Siggia, H. A. Rose, Phys. Rev. A {\bf 8} 423 -- 437 (1973).


\bibitem{Collins1984} J.~C.~Collins, {\it Renormalization: An Introduction to Renormalization, the Renormalization Group and the Operator-Product Expansion}, Cambridge Monographs on Mathematical Physics,
Cambridge University Press, Cambridge (1984).

\bibitem{Binoth2000} T. Binoth, G. Heinrich, Nucl. Phys. B {\bf 585}, 741 (2000). 

\bibitem{Hahn2005} T. Hahn, Comput. Phys. Commun. {\bf 168}, 78 (2005).

\end{thebibliography}
\end{document}